\documentclass[twoside]{article}
\usepackage{fleqn,espcrc2}

\usepackage{epsfig}
\usepackage{graphicx}
\usepackage[figuresright]{rotating}

\newcommand{\be}{\begin{equation}}
\newcommand{\ee}{\end{equation}}
\newcommand{\ba}{\begin{eqnarray}}
\newcommand{\ea}{\end{eqnarray}}

\newcommand{\AmS}{{\protect\the\textfont2
  A\kern-.1667em\lower.5ex\hbox{M}\kern-.125emS}}
\def\spose#1{\hbox to 0pt{#1\hss}}
\def\ltapprox{\mathrel{\spose{\lower 3pt\hbox{$\mathchar"218$}}
 \raise 2.0pt\hbox{$\mathchar"13C$}}}

\def\v{\vec}

\title{Specific heat and energy for the three-dimensional $O(2)$ model
       \thanks{Talk given by S. Holtmann, this work was supported by DFG Grant
       No. FOR 339/1-2.} }

\author{S. Holtmann with J. Engels and T. Schulze\\\vspace{0.5cm}
       {Fakult\"at f\"ur Physik, Universit\"at Bielefeld, \\
        D-33615 Bielefeld, Germany}}
        
\begin{document}
\begin{abstract}
We investigate the three-dimensional $O(2)$ model on lattices of size $8^3$ to
$160^3$ close to the critical point at zero magnetic field. We confirm
explicitly the value of the critical coupling $J_c$ found by Ballesteros et
al. and estimate there the universal values of $g_r$ and $\xi/L$. 
At the critical point we study
the finite size dependencies of the energy density $\epsilon$ and the specific
heat $C$. We find that the nonsingular part of the specific heat $C_{ns}$ is
linearly dependent on $1/\alpha$. From the critical behaviour of the specific
heat for $T\neq T_c$ on the largest lattices we determine the universal
amplitude ratio $A^+/A^-$. The $\alpha$-dependence of this ratio is close to the
phenomenological relation $A^+/A^-=1-4\alpha$.
\vspace{1pc}
\end{abstract}
\maketitle
\section{INTRODUCTION}
$O(N)$ models in three dimensions play an important part in condensed matter
physics and in quantum field theory, because many systems belong to the
corresponding universality classes. In three dimensions the case $N=2$ is a
special one, because the $O(2)$ model is the first $O(N)$ model (with increasing
$N$) exhibiting massless Goldstone modes. Furthermore its critical exponent
$\alpha$, which controls the critical behaviour of the specific heat, is
negative and very close to zero. In the famous shuttle experiment \cite{Shuttle}
the universal ratio $A^+/A^-$ and $\alpha$ have been determined
experimentally. Here we want to calculate this ratio from Monte Carlo
simulations.\\
\indent The $O(2)$-invariant nonlinear $\sigma$-model (or $XY$ model) for zero magnetic
field, which we examine here, is defined by the partition function
\be
Z\;=\;\int [d\v{\phi}] \exp\, [J \sum\limits_{<i,j>}
{\v{\phi}}_i \cdot {\v{\phi}}_j].
\ee
Here ${\v{\phi}}_x$ is a 2-component unit vector at site $x$ and $J=1/T$ is the
inverse temperature. We use the lattice average of the spins
\be
{\vec m}\;=\;{1\over V}\sum\limits_{i} {\v{\phi}}_i\;\;\;\mbox{with}\;\;\;V=L^3
\ee
to define the magnetization $M$, the order parameter of the system,
by
\be
M\;=\;\left<|{\vec m}|\right>,
\ee
the susceptibility $\chi$
\be
\chi\;=\;V\left<{\vec m}^2\right>,
\ee
and the Binder cumulant $g_r$
\be
g_r\;=\;{\left<({\vec m}^2)^2\right> \over {\left<{\vec m}^2\right>^2}}\,-\,3.
\ee
The second moment correlation length $\xi_{2nd}$ is calculated from
\be
\xi_{2nd}\;=\;\left( \frac{\chi/F-1}{4 \sin ^2(\pi/L)}  \right)^{1/2},
\ee
where $F$ is the Fourier transform of the correlation function at momentum
$p_\mu=2\pi \hat{e}_\mu/L$. Important observables for this work are the energy
$E$, the energy density $\epsilon$
\be
E\;=\;-\sum\limits_{<i,j>}{\v{\phi}}_i \cdot {\v{\phi}}_j\;\;\;,\;\;\;\epsilon\;=\;{\left<E\right> \over V}
\ee
and the specific heat $C$
\be
C\;=\;{\partial {\epsilon} \over \partial T}\;=\;{J^2 \over V} (\left<E^2\right>-\left<E\right>^2).
\ee
At the critical coupling $J_c$ the finite size behaviour of the energy density is
\be
\epsilon(L)\;=\;\epsilon_{ns}+q_{\epsilon}\,L^{{\alpha-1} \over \nu}
\label{epsilonL},
\ee
and that of the specific heat
\be
C(L)\;=\;C_{ns}+q_{c}\,L^{\alpha \over \nu}\,(1+q_{1c}L^{-\omega}).
\label{CL}
\ee
$\epsilon_{ns}$ and $C_{ns}$ are the nonsingular parts of the energy
density and of the specific heat, respectively. In the thermodynamic limit the
critical behaviour of $C$ for $T$ close to $T_c$ is
\be
C(t)\,=\, C_{ns}+{A^\pm \over \alpha}\,|t|^{-\alpha}\,[1+c_1^\pm
\,|t|^{\omega\nu}+c_2^\pm\,t],
\label{Cinv}
\ee
where $t=(T-T_c)/T_c$ is the reduced temperature. Here correction to scaling
terms have been included. We use the first irrelevant exponent
$\omega=0.79(2)$ from \cite{Hasenbusch} in the following fits.
\section{NUMERICAL RESULTS}
\label{section:results}
Our simulations were done on three-di\-men\-sio\-nal lattices with periodic 
boundary conditions and linear extensions $L=8-160$ using the Wolff 
single-cluster algorithm.

First we have determined again the critical coupling
$J_c$, utilizing the fact that Binder's cumulant should be finite size 
indepen\-dent at
criticality. We have interpo\-la\-ted data from simulations on lattices with
$L=24,36,48,72$ and $96$. The curves for the different lattices do not cross in
a single point due to small corrections to scaling. Including these we find that
the shift $\Delta$ from criticality of the crossing point between two lattices
of size $L$ and $L'$ ($b=L'/L$) is
\be
\Delta J_c^{L,L'}\;\propto\; \frac{1-b^{-\omega}}{b^{1/\nu}-1}L^{-\omega-1/\nu}.
\ee
The $\nu$-dependence of $\Delta J_c^{L,L'}$ is not relevant as long as $\nu \in
[0.669;0.675]$. Extrapolating to $L \to \infty$ we find the value
\be
J_c=0.454167(4),
\ee
in agreement with $J_c=0.454165(4)$ from \cite{Ballesteros}. In a similar way we
determine the universal values of $g_r$ and $\xi_{2nd}/L$ at $J_c$ to
\be
g_r=-1.758(2)\;\;\;\mbox{and}\;\;\;\xi_{2nd}/L=0.593(2),
\ee
in accordance with \cite{Pelissetto} and \cite{Hasenbusch}.\\
\indent In order to estimate the nonsingular parts $\epsilon_{ns}$ and $C_{ns}$
of the energy density and the specific heat the finite size effects of these
observables at $T_c$ were studied. The model was simulated at $J_c=0.454165$ on
lattices with $L=8$ to $L=160$. Fits to the data with Eqs.\ (\ref{epsilonL}) and
(\ref{CL}) show no corrections to scaling in case of the energy density and
only small corrections for the specific heat.$\!\!\!$
\vspace{-0.8cm}
\begin{figure}[ht!]
\begin{center}
   \epsfig{bbllx=100,bblly=264,bburx=425,bbury=587,
        file=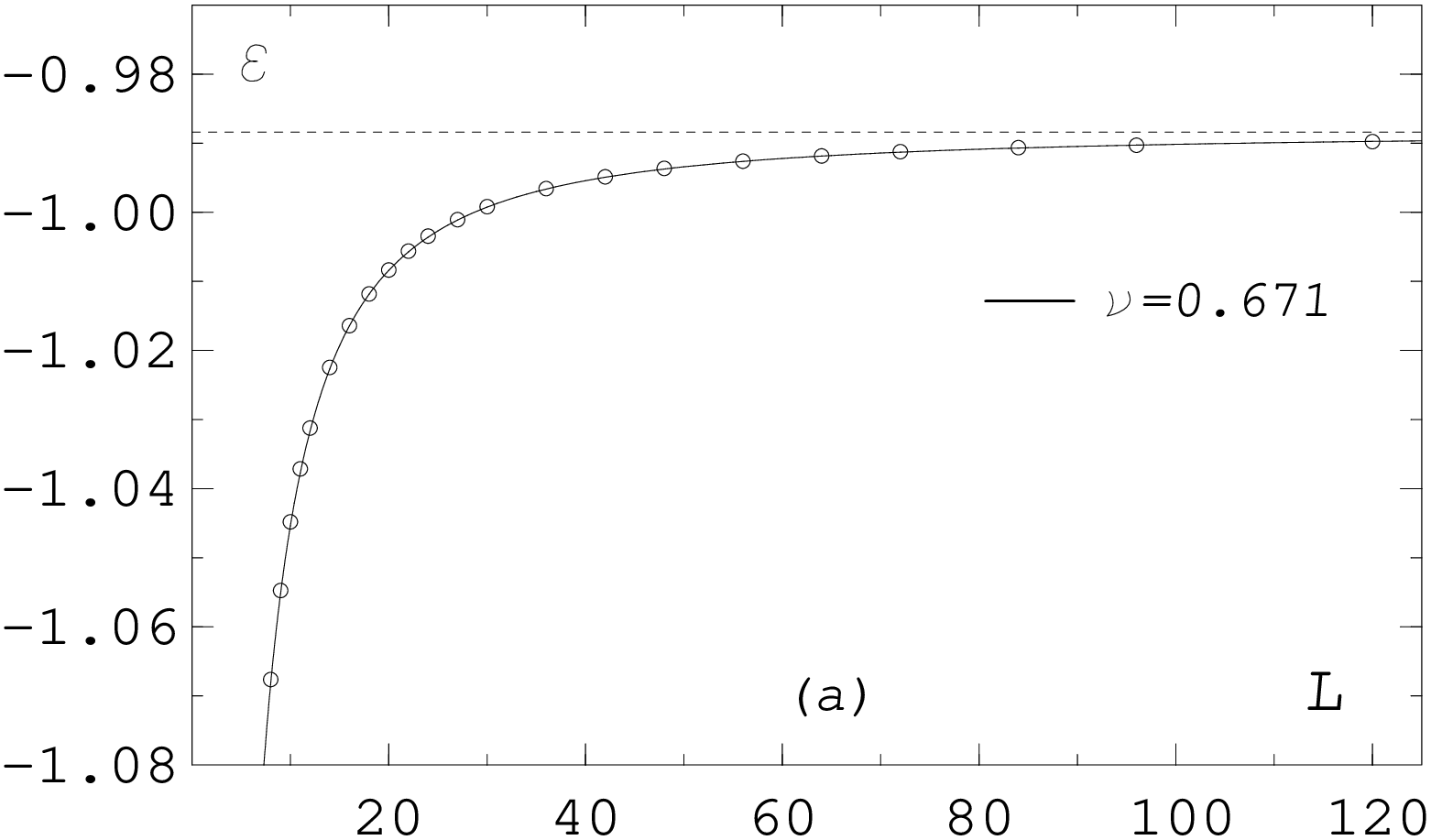,width=46mm}
   \epsfig{bbllx=100,bblly=264,bburx=425,bbury=587,
        file=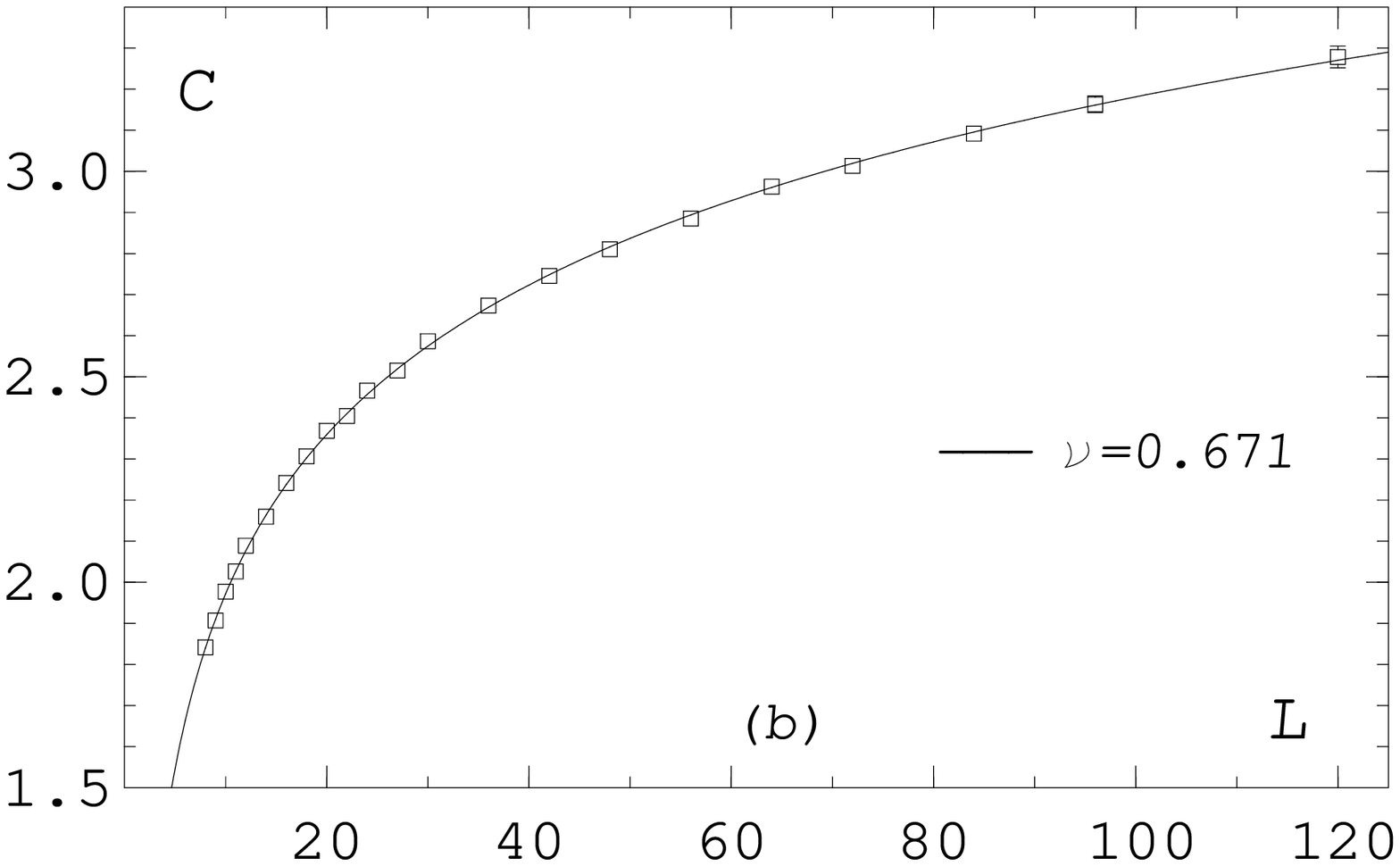,width=46mm}
\end{center}
\vspace{-0.9cm}
\caption{The energy density (a) and the specific heat (b) vs. $L$ at
  criticality. The dashed line shows $\epsilon_{ns}$ and the solid lines fits for
  $\nu=0.671$.}
\label{fig1} 
\vspace{-0.6cm}
\end{figure}
The quantity $\epsilon_{ns}$ exhibits no noticeable dependency on $\nu$, and we
find $\epsilon_{ns}=-0.98841(3)$. When we treat $\nu$ as a free fit parameter,
we get $0.671(2)$. In case of the specific heat the situation is
different. Its nonsingular part varies from about $50$ for $\nu=0.669$ to $16$
at $\nu=0.675$. This is so because the exponent $\alpha/\nu$ in Eq.\ (\ref{CL})
is approximately zero. As shown in Fig.\ \ref{fig2} the nonsingular part of the
specific heat $C_{ns}$ is linearly dependent on $1/\alpha$. We find
\be
C_{ns}(\alpha)\;=\;3.35(26)\,-\,{0.3176(43)\over \alpha}.
\label{cns}
\ee
\begin{figure}[ht!]
\begin{center}
   \epsfig{bbllx=120,bblly=264,bburx=425,bbury=560,
        file=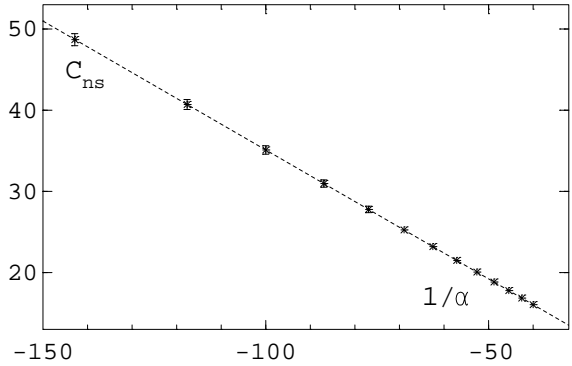,width=47mm}
\end{center}
\vspace{-0.9cm}
\caption{The nonsingular part of the specific heat. The dashed line is the
fit (\ref{cns}).} 
\label{fig2}
\vspace{-0.6cm}
\end{figure}
This result for $C_{ns}$ is used to analyse the specific heat at $T\neq T_c$
and $V\to \infty$. Our data are shown in Fig.\ \ref{fig3}. We have interpolated
the data from the largest lattices by reweighting. The resulting curves have
been fitted to the form (\ref{Cinv}). In the broken phase the correction to
scaling terms in the bracket are relevant for the fit, whereas these terms
are negligible for $T>T_c$. Again we observe an $\alpha$-dependency in the
fit parameters, especially for the amplitudes $A^\pm$:
\ba
A^+&=&0.3177(2)-3.29(4)\alpha+18.8(1.3)\alpha^2\\
A^-&=&0.3176(3)-1.97(4)\alpha+7.8(1.4)\alpha^2.
\ea
As expected, we obtain the same amplitudes for $\alpha \to 0$, and the
$1/\alpha$-pole term in $C_{ns}$ is cancelled exactly there: the same specific
heat data can as well be fitted at $\alpha=0$.\\
\indent The universal ratio $A^+/A^-$ can now be written as
\be
A^+/A^-\;=\;1-4.23(3)\alpha+3.3(1.8)\alpha^2+...\,.
\label{apam}
\ee
It is shown as solid line in Fig.\ \ref{fig4}. This result
is well in accordance with former results, e.g. from the
shuttle experiment \cite{Shuttle} and analytic determinations \cite{Pelissetto}
and \cite{Dohm}. The leading part is also close to the phenomenological relation
\cite{Hohenberg}
\be
A^+/A^-=1-4 \alpha~.
\ee
\begin{figure}[ht!]
\begin{center}
   \epsfig{bbllx=120,bblly=264,bburx=430,bbury=587,
        file=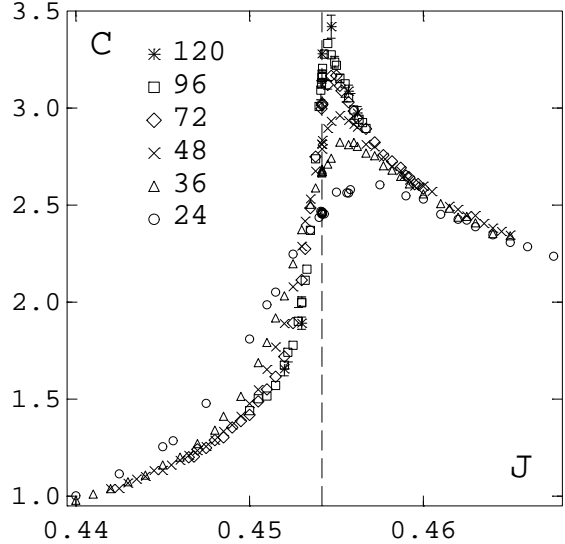,width=63mm}
\end{center}
\vspace{-0.9cm}
\caption{The specific heat for different $L$ vs. the coupling $J$.
The line shows the position of $J_c$.} 
\label{fig3}
\end{figure}

\begin{figure}[ht!]
\begin{center}
   \epsfig{bbllx=108,bblly=264,bburx=430,bbury=530,
        file=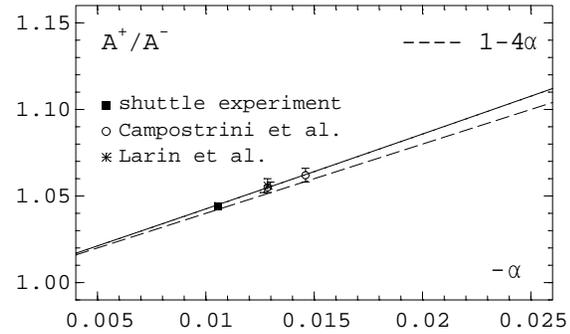,width=45mm}
\end{center}
\vspace{-0.9cm}
\caption{The universal ratio $A^+/A^-$ as a function of $-\alpha$.
The solid line shows the result (\ref{apam}).} 
\label{fig4}
\vspace{-0.6cm}
\end{figure}
\vspace{1cm}

\end{document}